\documentclass[prb,twocolumn,groupedaddress]{revtex4}
\usepackage{amsmath} 
\usepackage{amssymb} 
\usepackage{amsfonts}
\usepackage[dvips]{graphicx} 
\usepackage[]{epsfig} 
\begin{document}

\title{Substrate concentration dependence of the diffusion-controlled steady-state rate constant}
\author{J. Dzubiella} \email[e-mail address:] {jdzubiella@ucsd.edu}
\affiliation{NSF Center for Theoretical Biological Physics (CTBP),}
\affiliation{Department of Chemistry and Biochemistry, University of California, San Diego, La Jolla, California 92093-0365}
\author{J.~A. McCammon} 
\affiliation{NSF Center for Theoretical Biological Physics (CTBP),}
\affiliation{Department of Chemistry and Biochemistry, University of California, San Diego, La Jolla, California 92093-0365}

\date{\today}

\begin{abstract}

The Smoluchowski approach to diffusion-controlled reactions is
generalized to interacting substrate particles by including the
osmotic pressure and hydrodynamic interactions of the nonideal particles
in the Smoluchoswki equation within a local-density approximation.
By solving the strictly linearized equation for the time-independent case
with absorbing boundary conditions, we present an analytic expression for the
diffusion-limited steady-state rate constant for small substrate
concentrations in terms of an {\it effective} second virial
coefficient $B_2^*$. Comparisons to
 Brownian dynamics simulations
excluding HI show excellent agreement up to bulk number densities of
$B_2^* \rho_0 \lesssim 0.4$ for hard sphere and repulsive Yukawa-like
interactions between the substrates. Our study provides an alternative
way to determine the second virial coefficient of interacting
macromolecules experimentally by measuring their steady-state rate
constant in diffusion-controlled reactions at low densities.
\end{abstract}

%\pacs{61.20.Ja, 68.08.Bc, 87.16.Ac, 87.16.Uv}
\maketitle
\section{Introduction}

Simple irreversible reactions of the type A+B$\rightarrow$ A (with A
the sink and B the substrates) are commonly found in many
(bio)chemical processes, such as fluorescence quenching, enzyme
catalysis, polymerization, or colloid and protein aggregation, just
to mention a few examples.\cite{rice} The key parameter for these
processes is the reaction rate constant, a measure for the number of
reactions per unit time, first addressed in the pioneering and now
classical works of Smoluchowski\cite{smoluchowski} and Debye
\cite{debye} decades ago.  Since then, various improvements and
refinements have been made in predicting the rate constants for
diverse
reactions,\cite{felderhof:jcp:1976,northrup:jcp:1979,northrup:jcp:1984,calef:annrev:1983,rice,berg:annrev:1985,dong:jcp:1989,zhou:jcp:1991,ibuki:jcp:1997,yang:jcp:1999}
in particular, including solvent-mediated hydrodynamic interactions
(HI) between sink (A) and substrate (B),\cite{calef:annrev:1983} or
examining effects of a non-zero concentration of the sink
particles.\cite{felderhof:jcp:1976,szabo:prl:1988,gopich:jcp:2002} In
most of the previous studies, interactions between the substrate
particles were ignored, which is only justified in the case of very
weakly interacting substrate particles or at infinite dilution. In
recent attempts the influence of the excluded volume of the substrate
particles was examined\cite{jung:jpc:1997,lee:jcp:2000} and predicted
to yield an increased reaction rate with increasing excluded volume or
substrate concentration. Along these lines, Senapati and
McCammon\cite{senapati:jcp:2004} also gave evidence for a strong
influence of substrate interactions (without HI) on the reaction rate
by means of Brownian Dynamics (BD) computer simulations. We tie up to
these studies in this work and aim at a systematic examination of the
substrate concentration-dependence of the rate constant, while the
sinks remain at infinite dilution.

For many of the reactions mentioned above the rate-limiting factor is
the diffusional encounter of the reactants, particularly when the
subsequent transformation does not involve a large activation barrier.
In our work we will focus on the steady-state (long-time) rate of
these diffusion-limited or diffusion-controlled reactions. The usual
theoretical approach is based on the Smoluchowski equation (SE) which
becomes time-independent in the steady-state.  For weakly interacting
particles or for small concentrations, the inhomogeneous density
profile of the substrates around the sink varies smoothly over
distances larger compared to the typical interaction range and well
established generalizations of the SE are available,\cite{dhontbook}
often used for instance for the problem of colloidal
sedimentation. The generalized SE employs the osmotic pressure and
density-dependent mobility of the interacting particles within a
local-density approximation (LDA)\cite{hansen:mcdonald} which assumes
local homogeneity of the density and is justified whenever the density
varies slowly in space. Using a simple model with a spherical,
isotropically reactive sink particle, we will show that a strictly
linearized version of the generalized SE allows us to write down an
analytic solution for the first order correction of the rate constant
in linear order of substrate concentration.  As a result, the
correction coefficient is basically given by the second virial
coefficient $B_2$, the first order correction coefficient in the
expansion of the virial equation of state, but must be corrected for
HI and is also influenced by the interaction between sink and
substrate. Fortunately, most chemical reactions occur at small
densities of the reactants so that our result should be valid for a
wide range of processes and systems. Despite the simplicity of our
model the results are general and should be applicable to more
realistic and complicated systems.

In
order to examine the range of validity of our theoretical result we
perform Brownian dynamics (BD) simulations for different systems in
which the interaction between the B particles is varied from hard
sphere to attractive and repulsive Yukawa-like interactions, the
latter typically found in ionic solutions. The BD simulations do not
include HI, which can be accurately treated only by more
sophisticated and computationally more expensive means, such as
Lattice-Boltzmann methods\cite{lb} or others.\cite{kapral,tucci:jcp:2004,tanaka:prl:2000} However, in
comparison to the theory excluding HI we find excellent agreement up
to substrate densities $B_2\rho_0\simeq 0.4$, showing the reliability
of our theoretical concept, and proving its applicability to the case
when HI can be neglected, i.e. for long-ranged interactions. In
principle, our study opens up an alternative way to determine the
second virial coefficient of proteins or other particles by measuring
the reaction rate in diffusion-limited reactions at small substrate
densities.

The paper is organized as follows: In section II we present the basic
equations of motion and approximations of our theory and arrive at a
first order linear differential equation which can be 
solved analytically. The solution for the steady-state rate constant and
density profile are presented and discussed in section III. A
systematic comparison to BD simulations in order to examine the range
of validity of our theory follows in section IV. Sec. V concludes our
work with a few final remarks.
\section{Generalized Smoluchowski theory}
\subsection{Model and basic equations}
Let us consider a spherical, isotropically reactive sink particle A
with diameter $\sigma_{\rm AA}$ which reacts with a B particle
(substrate) of size $\sigma_{{\rm BB}}=:\sigma$ when they touch at a
center-to-center distance $r=\sigma_{\rm AB}=(\sigma_{\rm
AA}+\sigma)/2$. The sinks are at infinite dilution while the B
particles have a bulk number density $\rho_0$.  Furthermore a
potential $V_{\rm AB}(r)$ is acting between the sink and the substrate 
particles. Both A and B are dispersed in a solvent, which is taken into
account by a position-dependent diffusion constant $D_0({\bf r})$ of the substrates,
reflecting the diffusion of an isolated B particle relative to the A
particle.  We assume the sink to be completely at rest, which is
justified for large sinks $\sigma_{\rm AA}\gg\sigma$. For
close distances to A, $D_0({\bf r})$ is determined by the HI between A and B, while for large distances when HI
are negligible, we obtain $D_0$ by the Stokes-Einstein relation
\begin{eqnarray}
D_0(|{\bf r}|\rightarrow\infty)=D_0=(3\beta\pi\eta\sigma)^{-1},
\label{eq:stokes}
\end{eqnarray}
where $\beta^{-1}=k_BT$ is the thermal energy, and $\eta$ the solvent
viscosity. 

The time-dependent Smoluchowski equation (SE) for
noninteracting particles moving in a potential $V_{\rm AB}({\bf r})$
with a position-dependent diffusion constant $D_0({\bf r})$ is given by
\begin{eqnarray}
\partial \rho({\bf r},t)/\partial t & = & -{\bf \nabla}\cdot {\bf J}({\bf
r},t) \\ & = & {\bf \nabla}\cdot D_0({\bf r})\left[\rho({\bf r},t)\frac{\beta \partial
 V_{\rm AB}({\bf r})}{\partial r}+{\bf \nabla}\rho({\bf r},t)\right] \nonumber.
\label{eq:se}
\end{eqnarray}
For weak density inhomogeneities, which is always true for small
densities or weak interactions, one can introduce substrate
interactions by using the generalized Stokes-Einstein relation
\cite{dhontbook}
\begin{eqnarray}
D(\rho)=M(\rho)\frac{{\rm d}\Pi(\rho)}{{\rm d}\rho},
\label{eq:gstokes}
\end{eqnarray}
where $D(\rho)$ is the collective diffusion coefficient, $M(\rho)$ is
the density dependent mobility of the substrate particles, and
$\Pi(\rho)$ the osmotic pressure of the interacting particles. This is
a generalized Stokes-Einstein equation in the sense that it
generalizes (1) to the case of interacting particles in a homogeneous
solution. The mobility $M(\rho)$ is a reciprocal friction and is
defined as the proportionality constant between the drift velocity and
total force on a Brownian particle
($velocity\;=\;M(\rho)\;\times\;total\;force$) in a steady-state
situation. Within the LDA the generalized Stokes-Einstein relation is
applied to the local, position-dependent density $\rho({\bf r})$ of
the substrate particles, which we assume to vary smootly over
distances larger compared to the range of the interaction potential.
The equation of motion can now be written
as\cite{dhontbook}
\begin{eqnarray}
&&\partial\rho({\bf r},t)/\partial t = -{\bf \nabla}\cdot{\bf J}({\bf
r},t)\label{eq:gse} \\ & = & {\bf \nabla}\cdot M(\rho({\bf
r},t),{\bf r})\left[\rho({\bf r},t)\frac{\partial V_{\rm AB}({\bf
r})}{\partial r}+[{\bf \nabla}\rho({\bf r},t)]\frac{{\rm
d}\Pi(\rho({\bf r},t)}{{\rm d}\rho({\bf r})}\right], \nonumber
\end{eqnarray}
and is a generalization of Eq.~(\ref{eq:se}) to weakly interacting
substrate particles. The last term on the rhs of Eq.~(\ref{eq:gse})
accounts for the force on a Brownian particle due to an unbalanced
osmotic pressure caused by a concentration gradient of interacting
particles. In addition, the density-dependent mobility corrects for HI
between B particles. Note that for ideal particles ${\rm d}\Pi/{\rm
d}\rho=k_BT$, and with no HI between B particles $M(\rho({\bf r}),{\bf
r})=\beta D_0({\bf r})$ and we find (2) again. To account for
reactions at the sink we have to solve Eq. (\ref{eq:gse}) with the
boundary condition
\begin{eqnarray}
4\pi\sigma_{\rm AB}^2 ({\bf r}/r) {\bf J} ({\bf r},t)|_{\sigma_{\rm AB}} = -k_i \rho({\bf r},t)|_{\sigma_{\rm AB}},
\end{eqnarray}
where $k_i$ is the intrinsic rate constant. In the limit of
fully absorbing boundary conditions, $k_i \rightarrow \infty$, while
$\rho(\sigma_{\rm AB})\rightarrow 0$, and the reaction rate is only
limited by the diffusion of the approaching substrates.  We are
interested in the steady-state solution for the spherically symmetric
problem, with which the equation of motion reduces to 
\begin{eqnarray}
0 = \frac{\partial}{\partial r }\left(r^2
M(\rho(r),r)\left[\rho(r)\frac{\partial V_{\rm AB}(r)}{\partial
r}+[\frac{\partial}{\partial r}\rho(r)]\frac{{\rm d}\Pi(\rho(r))}{{\rm
d}\rho(r)}\right]\right)
\label{eq:gse2}
\end{eqnarray}
with the boundary conditions for diffusion-controlled reactions
\begin{eqnarray}
4\pi\sigma_{\rm AB}^2 j(\sigma_{\rm AB}) =  k \rho_0,
\label{eq:bc}
\end{eqnarray}
and 
\begin{eqnarray}
\rho(\sigma_{\rm AB})=0\;\;\;{\rm and}\;\;\rho(\infty)=\rho_0,
\label{eq:bc2}
\end{eqnarray}
where $j(\sigma_{\rm AB})$ is the absolute flux perpendicular through the sink
surface at radial distance $\sigma_{\rm AB}$ from the sink center, and
the quantity $k$ is the diffusion-controlled steady-state rate
constant which we are interested in. 

The osmotic pressure can be
expressed in terms of the equation of state, or in general, the virial
equation, which is an expansion of the pressure in terms of the
particle density $\rho$. For small densities the series is usually
truncated after the second order and the osmotic pressure reads
\begin{eqnarray}
\beta\Pi(\rho) = \rho + B_2\rho^2.
\label{eq:press}
\end{eqnarray}
Conveniently, the second virial coefficient $B_2$ can be calculated
from the pair interaction $V_{\rm BB}({\bf r})=:V({\bf r})$ between
the substrates via\cite{hansen:mcdonald}
\begin{eqnarray}
B_2 =  \frac{1}{2} \int {\rm d}{\bf r} [1-\exp(-\beta V({\bf r}))].
\end{eqnarray}
The local mobility of the substrates will be in general reduced by a 
nonzero concentration of other surrounding particles due to HI, 
and for small densities the correction is linear in $\rho$, 
so that we can write 
\begin{eqnarray}
M(\rho({r}),{ r}) = \beta D_0({ r})\left[1-\alpha\rho({ r})\right],
\label{eq:m}
\end{eqnarray} 
where $\alpha>0$ is the first order correction coefficient. In
(\ref{eq:m}) we assume that the HI between A and B, expressed in
$D_0(r)$, are independent of the HI between B particles and the
resulting mobility can be factorized. For the case of hard spheres
with diameter $\sigma$ it is well established that $\alpha\sigma^{-3}
\simeq 3.43$.\cite{dhontbook} Substituting (\ref{eq:m}) and the
derivative of (\ref{eq:press}) with respect to $\rho$ into
Eq. (\ref{eq:gse2}), integrating and strictly linearizing in $\rho_0$ we
find
\begin{eqnarray}
\frac{C}{D_0(r)r^2} = \rho(r)\left[\frac{\beta \partial V_{\rm AB}(r)}{\partial
r}+\frac{2B_2^*C}{D_0(r)r^2}\right]- \frac{\partial}{\partial r}\rho(r),
\label{eq:dg}
\end{eqnarray}
where we introduced the {\it effective} second virial coefficient
\begin{eqnarray}
B_2^*=B_2-\alpha/2,
\label{eq:b2}
\end{eqnarray}
and which has to be solved with boundary conditions Eqs. (\ref{eq:bc})
and (\ref{eq:bc2}). The equilibrium pair interaction, described by $B_2$,
and the dynamic interaction, absorbed in $\alpha$, compete likewise
now in altering the rate of the reaction, and can hence be summarized in 
the single parameter $B_2^*$ given in (\ref{eq:b2}).
This interesting feature will be discussed in more detail in
the next section, where the result for $k$ is
presented. Eq. (\ref{eq:dg}) is a first order linear differential
equation and can be solved analytically. The parameter $C$ is an
integration constant, determined by the boundary conditions, and is
related to the rate constant with $C=\rho_0k/4\pi$.

\subsection{Result for the steady-state rate constant}
From the linearized solution of (\ref{eq:dg}) we arrive at the result
for the diffusion-controlled steady-state rate constant valid for
small densities (or weak interactions)
\begin{eqnarray}
k = k_D \left[1+B_2^*\rho_0 \frac{I(\infty)}{I_1(\infty)} \right],
\label{eq:k}
\end{eqnarray}
where $k_D$ is the classical result of Debye \cite{debye} for ideal or infinitely diluted substrates in a nonzero potential $V_{\rm AB}(r)$ 
\begin{eqnarray}
k_D =  \left\{\int_{\sigma_{\rm AB}}^\infty {\rm d}r\frac{\exp[\beta V_{\rm AB}(r)]}{4\pi D_0(r)r^2}\right\}^{-1}=\{I_1(\infty)\}^{-1}
\label{eq:kD}
\end{eqnarray}
and 
\begin{eqnarray}
I(r)=2\frac{I_0(r)I_1(r)-I_2(r)}{I_1(\infty)}
\end{eqnarray}
with the integrals
\begin{eqnarray}
I_0(r)=\int_{\sigma_{\rm AB}}^r {\rm d}r'\; \frac{1}{4\pi D_0(r')r'^2},
\end{eqnarray}
\begin{eqnarray}
I_1(r) = \int_{\sigma_{\rm AB}}^r {\rm d}r'\;\frac{\exp[\beta V_{\rm AB}(r')]}{4\pi D_0(r')r'^2},
\end{eqnarray}
and 
\begin{eqnarray}
I_2(r) = \int_{\sigma_{\rm AB}}^r {\rm d}r'\;\frac{\exp[\beta V_{\rm AB}(r')]\;I_0(r')}{4\pi D_0(r')r'^2},
\end{eqnarray}
which are functions depending on the particular interaction $V_{\rm
AB}(r)$ and the position-dependent diffusion constant $D_0(r)$.  When interactions
between A and B can be neglected, namely $D_0(r)=D_0$, and $V_{\rm
AB}(r)=0$, it follows $I(\infty)/I_1(\infty)=1$, and the result for
the steady-state rate constant reduces to
\begin{eqnarray}
k = 4\pi D_0\sigma_{\rm AB} (1+B_2^*\rho_0),
\end{eqnarray}
which for noninteracting particles or at infinite dilution of
species B gives the classical Smoluchowski result
\cite{smoluchowski}
\begin{eqnarray}
k_0 = 4\pi D_0\sigma_{\rm AB}.
\label{eq:k0}
\end{eqnarray}
The constant $I(\infty)/I_1(\infty)$ in (\ref{eq:k}) depends on the
particular interaction $V_{\rm AB}(r)$ and $D_0(r)$ between A and B
particles, but is always close to $1$ for Lennard-Jones or Yukawa like
interactions typically found in solutions, and a $D_0(r)$ for instance
given by the Oseen tensor.\cite{northrup:jcp:1984} The major
contribution to the density correction stems from the effective second
virial coefficient (\ref{eq:b2}), which is a corrected second virial
coefficient for the dynamic situation and features the following
interesting trends for the steady-state rate constant. Generally
$\alpha>0$, imposing that HI will decrease the substrate
mobility. However, if the substrate-substrate interaction is strongly
repulsive, $B_2>\alpha/2$, so that $B_2^*$ is positive, an enhanced
reaction rate is predicted compared to the noninteracting case. On the
contrary, a mainly attractive pair interaction will in general lead to
a negative $B_2^*$ and decrease the reaction rate. For long-ranged
interactions $|B_2|$ may be an order of magnitude larger than
$\alpha/2$ and HI can be neglected, such that $B_2^*\simeq B_2$. When
the repulsive interaction leads to a $B_2$ comparable to $\alpha/2$,
both contributions can cancel each other and the reaction rate just
changes little or not at all with increasing (small) density or (weak)
interaction.  For instance, for the repulsive hard sphere case the
effective second virial coefficient $B_2^*\sigma^{-3}\simeq
2.09-3.43/2\simeq 0.38$ is small due to a slowing of the diffusion of the
substrates which is comparable to the effects of the enhanced osmotic
pressure in the system, leading just to a minor increase of the rate
constant with density.  Our theory is in agreement with previous
work\cite{jung:jpc:1997,lee:jcp:2000} where the excluded volume of the
substrates is predicted to lead to an enhanced reaction rate
(excluding HI). It is worth mentioning that $B_2^*$ is half of the
coefficient of the linear first order correction in $\rho_0$ of the
collective diffusion coefficient of the substrates in bulk solution,
latter evident from the generalized Stokes-Einstein relation
(\ref{eq:gstokes}). The range of validity of (\ref{eq:k}) will be
examined in section III for hard sphere and repulsive and attractive
Yukawa interactions excluding HI.

\subsection{Result for the density profile}

The general solution for the diffusion-controlled steady-state density profile in linear order in $\rho_0$ reads
\begin{eqnarray}
  \rho(r)=\rho_0\frac{ k_{\rm D} I_1(r)}{\exp[\beta V_{\rm
  AB}(r)]}\left(1+B_2^*\rho_0\left[\frac{I(\infty)}{I_1(\infty)}-\frac{I(r)}{I_1(r)}\right]\right).
\end{eqnarray}
In order to better identify how the substrate interactions affect the density profiles 
we rewrite the solution for the case of no interactions between substrate and 
sink, $D_0(r)=D_0$ and $V_{\rm AB}(r)=0$:
\begin{eqnarray}
\rho(r)=\rho_{\rm id}\left(1+B_2^*\rho_0\frac{\sigma_{\rm AB}}{r}\right),
\label{eq:dens}
\end{eqnarray} 
where 
\begin{eqnarray}
\rho_{\rm id}(r)=\rho_0\left(1-\frac{\sigma_{\rm AB}}{r}\right)
\label{eq:densid}
\end{eqnarray}
is the classical result for ideal or infinitely diluted substrate
particles when $V_{\rm AB}=0$ and $D_0(r)=D_0$.
\cite{smoluchowski,felderhof:jcp:1976} Eq. (\ref{eq:dens}) predicts
that a positive $B_2^*$ will enhance the substrate density close to
the absorbent due to a higher (positive) bulk pressure of the B
particles compared to the noninteracting case. This is opposite to the
case of negative $B_2^*$ (mainly attractive interactions) where a
depletion of B particles takes place close to A due to a negative bulk
pressure, and/or a larger immobility of the approaching
substrates. Comparing these trends to the rate constant (\ref{eq:k})
we conclude that an enhanced or lower density close to the sink
increases or decreases the reaction rate, respectively, at least for small
densities. These predictions will be examined in the next section with
BD simulations.

\section{Brownian dynamics simulation}
\subsection{Simulation details}
For a verification of the theory in the previous section we perform
standard Brownian Dynamics simulations using the integration technique
of Ermak and McCammon,\cite{ermak} and in addition accounting for hard
sphere overlap.\cite{cichocki} In order to conduct a clear and
transparent comparison to the  theoretical result we neglect HI in
the BD simulations. An accurate treatment of HI is only given by more
sophisticated and computationally expensive techniques, such as
Lattice-Boltzmann\cite{lb} or others.\cite{kapral,tucci:jcp:2004,tanaka:prl:2000} Thus, in the following
$D_0(r)=D_0$ given by (\ref{eq:stokes}), and $B_2^*=B_2$.  In the BD simulations, the
steady-state rate constant is determined by calculating the survival
probability $S(t)$ of the sink particle A  in the steady-state, which
is given by
\begin{eqnarray}
S(t)=\exp(-k\rho_0 t)
\label{eq:S}
\end{eqnarray}
in the presence of reactive B particles with bulk density $\rho_0$. In
our simulation model, the sink is fixed in the center of the
periodically repeated, cubic simulation box with length $L$. Substrate
(B) particles react with the sink (A) as soon they touch it, which
happens at a center-to-center distance $r_{\rm AB}=\sigma_{\rm
AB}$. The trajectory of a B particle is terminated after reaction.
The survival probability can be easily measured by forming an
histogram of the times between two reactions and normalize it with
respect to the number of total reactions in one simulation. To account
for the steady-state case and simulate a fixed density $\rho_0$ far
away from the sink, the annihilated B particle is reinserted randomly
at a position close ($r_{\rm AB}>0.45L$) to the box edges. Due to the
long range of the steady-state density-profile (\ref{eq:dens}), finite
size effects can be large for too small box sizes for all densities
and have been analyzed by performing finite size scaling
simulations. Eventually the box lengths used in our simulations range
from $L=50\sigma$ up to $L=200\sigma$ depending on density and
interaction of the B particles. This involved simulations of
$N_B=1000-30000$ substrate particles using $5\times10^7$  to
 $2\times10^5$ timesteps. The integration timestep was chosen to be
$0.003\tau_B$, where $\tau_B=\sigma^2/D_0$ is a typical Brownian
timescale in the simulation. Verlet-neighbor lists were used to
optimize computational time.\cite{allen:tildesley} Densities up to
$B_2\rho_0\lesssim 1$ could be simulated; for larger densities the
system size and statistical errors become too large for reasonable
output.
\subsection{Systems}
In order to perform a systematic comparison to
the theoretical results we consider four different systems,
I, II, III, and IV which differ in the interaction between A and B,
and B and B particles. In all systems the interactions are in general
given by
\begin{eqnarray}
\beta V_{ij}(r) = \beta V_{\rm HS}(r_{ij})+U_{ij}\frac{\sigma_{ij}}{r_{ij}}\exp[-\kappa(r_{ij}-\sigma_{ij})]
\label{eq:yuk}
\end{eqnarray}
where $\beta V_{\rm HS}$ is the hard sphere interaction
\begin{eqnarray}
\beta V_{\rm HS}= \begin{cases}
\infty & {\text {for}} \,\,r_{ij} \leq \sigma_{ij} ; \cr
   0 & {\rm else}, \cr
\end{cases}
\end{eqnarray}
and the second term in (\ref{eq:yuk}) is a Yukawa interaction with
energy scale $U_{ij}$, inverse screening length $\kappa$, and
$i,j=A,B$. $\kappa$ will be fixed to $\kappa\sigma=1$ in all our four
systems. We also fix the length scale to $\sigma_{\rm AB}=1.5\sigma$,
which is given when the sink particle is twice as large as the
substrate particle $\sigma_{\rm AA}=2\sigma$. We note already that we
have simulated selected densities for all four systems for two other
sink sizes, namely $\sigma_{\rm AB}=\sigma$ and $\sigma_{\rm
AB}=2\sigma$, and did not find any deviation from the results later in
the work. For high asymmetries $\sigma_{\rm AA}\gg\sigma$, which is the
more realistic case, the BD simulations unfortunately become
computationally expensive due too a large number of substrate particles
which have to be simulated.  Note again that the range of the
steady-state density profile increases essentially linearly with
$\sigma_{\rm AB}$, see Eq. (\ref{eq:dens}); however, we stress that
this does not affect our theoretical result (\ref{eq:k}) which should
be valid for all size ratios, and we have chosen smaller sink sizes
only for computational convenience.
 
In system I we simply consider $U_{ij}=0$, in which case all
interactions are hard sphere like. The second virial coefficient for
HS is $B_2\sigma^{-3}=2\pi/3\simeq 2.09$. In system II Yukawa-like
repulsion $U_{\rm BB}=1$ is added between the substrate particles. The
second virial coefficient increases to a total of
$B_2\sigma^{-3}\simeq 2\pi/3+11.23 \simeq 13.32$.  In system III a
Yukawa attraction is added between sink (A) and substrate (B)
particles, $U_{\rm AB}=-1$, while keeping the B-B interaction as in
system II. Finally, in system IV we choose $U_{\rm AB}=0$, no interaction
between A and B, but here we focus on attraction between the B
particles, $U_{\rm BB}=-0.5$, resulting in a total negative second virial
coefficient $B_2\sigma^{-3}\simeq-4.62$.  The system parameters and
corresponding values of the second virial coefficient and the constant
$I(\infty)/I_1(\infty)$ are summarized in Tab. 1.

\begin{table}
\begin{center}
\begin{tabular}{c | c  c  c c c}
   System & $U_{\rm AB}$ & $U_{\rm BB}$ & $\sigma_{\rm AB}/\sigma$ &$B_2\sigma^{-3}$ & $I(\infty)/I_1(\infty)$\\ 
\hline 
  I  & 0.0 & 0.0  & 1.5  & 2.09 & 1.00 \\
 II  & 0.0 & 1.0  & 1.5  & 13.32 & 1.00\\
 III & -1.0&  1.0  & 1.5 & 13.32 &1.11 \\  
IV & 0.0 & -0.5 & 1.5 & -4.62 & 1.00 \\
\end{tabular}
\caption{}
\label{tab1}
\end{center}
\end{table}

Typical examples of the calculated survival probabilities (\ref{eq:S}) are shown in
Fig.~1 for systems I, II, and IV on a logarithmic ordinate and show
linear behavior for all times within the uncertainty of the
statistics. The density for system I
(hard spheres) is chosen to be very small ($B_2\rho_0=0.001$) and
the survival probability matches with the classical theoretical result $S(t)=\exp(-k_0\rho_0
t)$, also shown in Fig.1 as dashed line. As expected from the theory, a
positive or negative $B_2$ (systems II and IV) increases or decreases
the rate constant, respectively, as is indicated by an increased or
decreased absolute slope of the $S(t)$ curves. Linear fits through these
curves determine our 'experimental' values of the rate constant for
all systems, while the regression coefficient of the fitting procedure 
provides the error bars.

\subsection{Results}

 \begin{figure}
 \begin{center}
    \epsfig{file=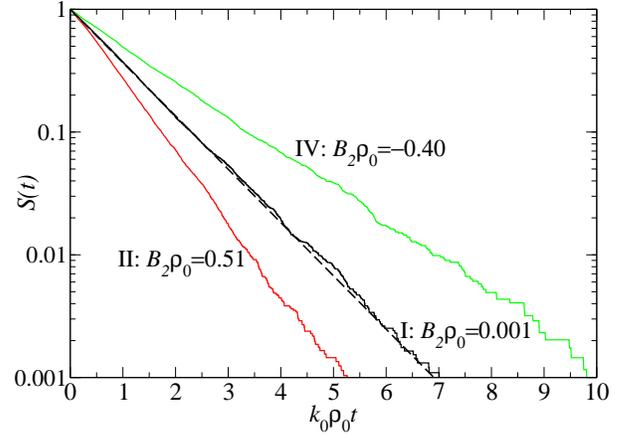, width=7cm, angle=-90}
  \caption{Typical examples for the survival probability S(t) on a
  logarithmic scale versus time scaled by the ideal substrate rate
  constant $k_0$ (\ref{eq:k0}) and substrate bulk density $\rho_0$. The straight
  dashed line is the ideal result $S(t)=\exp(-k_0\rho_0 t)$, while the
  noisy data are BD simulation results for chosen densities in systems
  I, II, and IV.}
 \label{fig:result}
 \end{center}
 \end{figure}

 \begin{figure}
 \begin{center}
    \epsfig{file=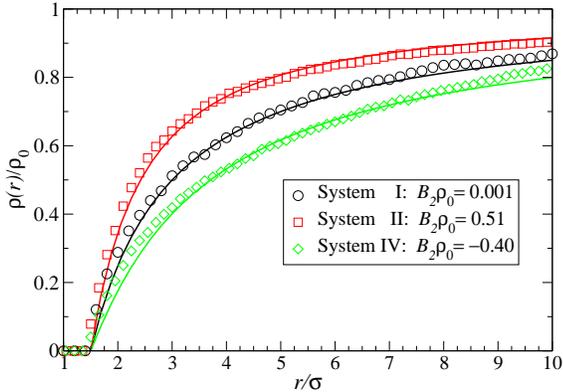, width=6.5cm, angle=-90}
  \caption{Steady-state density profiles $\rho(r)$ of the substrates
  around the sink particle from BD simulations (symbols) and theory
  (lines) according to Eq. (\ref{eq:dens}) for three different
  parameter sets. $r$ is the distance from the center of the sink. The
  circles are close to the infinite dilution limit (\ref{eq:densid}),
  while squares and diamonds are for positive and negative $B_2$ for
  nonzero bulk densities $\rho_0$, respectively.}
 \label{fig:result}
 \end{center}
 \end{figure}

 \begin{figure}
 \begin{center}
    \epsfig{file=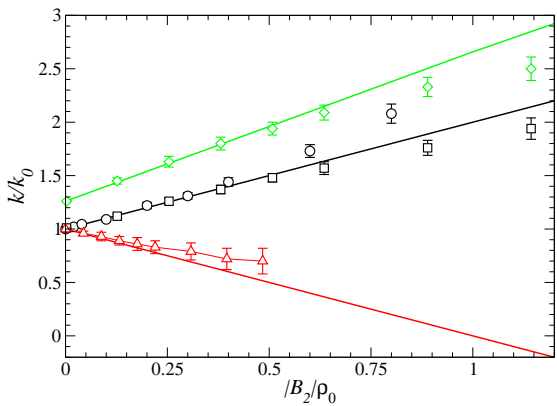, width=6.5cm, angle=-90}
  \caption{Dimensionless reaction rate $k/k_0$ from BD simulations
   (symbols) and theory (lines, Eq.(\ref{eq:k})) versus density of the
   substrate particles for the four different systems I (circles), II
   (squares), III (diamonds), and IV (triangles).  Note that the
   density is scaled by the corresponding second virial coefficient,
   see Tab. I, that is why the theoretical curves for I and II are on
   top of each other and have the same absolute slope as the curve for IV.}
 \label{fig:result}
 \end{center}
 \end{figure}

Examples of the density profiles for system I, II, and IV are shown in
Fig.~2 compared to the theoretical prediction (\ref{eq:dens}) and are
in very good agreement. The density for system I is in the very
dilute region ($B_2\rho_0=0.001$) and the profile matches the
$1-\sigma_{\rm AB}/r_{\rm AB}$ behavior (\ref{eq:densid}) for ideal substrate particles. Further on,
the predicted density increase (decrease) around the sink by repulsive
(attractive) substrate interactions is verified by the BD
simulations. All observed density profiles exhibit a very small but
nonvanishing density value at contact $\rho(\sigma_{\rm AB})\simeq 0.05$, showing
that the assumption $\rho(\sigma_{\rm AB})=0$ for absorbing boundary
conditions in the theory is justified, but not perfect. This contact value 
increases (decreases) slightly with positive (negative) $B_2$.

Results for the rate constant are shown in Fig. 3 where we plot the
steady-state rate constant scaled by the ideal rate constant $k_0$
given by (\ref{eq:k0}) versus the substrate density scaled by the
second virial coefficient $B_2$, which can be different for each
system, see Tab. 1. Note again that we neglect HI in our simulations,
such that $\alpha=0$ and $D_0(r)=D_0$, and $B_2^*=B_2$ in the
theory. We find excellent agreement to the theory (lines) within the
statistical uncertainties of the simulation for the systems with
repulsive substrate interaction (I-III) up to densities
$B_2\rho_0\simeq 0.4$.  For system I, the hard sphere case, the
nonlinear regime takes over for larger densities and the theoretical
prediction underestimates the rate constant. For systems II and III we
still find good agreement (theory within the error bars) till up to
$B_2\rho_0 \simeq 0.7$.  This agreement for a larger density range as
compared to HS could be anticipated since for soft repulsive
interactions it is known that higher order virial coefficients ($B_3$,
and so on) are less important even for higher densities.
\cite{likos:physrep:01} However, for larger densities the shortcomings
of the linearized Smoluchowski equation also take their effects.  In
the system with attractive substrate interactions (IV) the agreement
is good only for smaller densities ($B_2\rho_0\lesssim
0.20$). Equilibrium statistics has shown that for attractive
interactions, a LDA approach usually fails to give an accurate
description of the system behavior even for weak interactions, which
could also be the case here. Secondly, finite size effects in the
simulation are larger due to a longer ranged density profile
(\ref{eq:dens}) and cannot be excluded to explain the discrepancies
for larger densities. The theoretical curves for I and II are on top
of each other and have the same absolute slope as the data for IV because the
density is scaled by the corresponding $B_2$. For zero density the
curves for I, II, and IV intersect at $k/k_0=1$ which is the classical
ideal gas limit for no interactions between B and B, and A and
B. Curve III intersects at a larger value $k/k_0 \simeq 1.26$ due to the
additional A-B interaction in this system. This is also the main
reason for the increased absolute slope compared to I, II, and IV, but
additionally the constant $I(\infty)/I_1(\infty) \simeq 1.1$ is enhanced by
the attractive A-B interaction in this system.

\section{Concluding remarks}
In conclusion we have derived an analytic expression for the density
profile and rate constant for weakly interacting substrate particles
for the steady-state case of diffusion controlled reactions. For this purpose
we used the Smoluchowski equation, which was generalized within a LDA 
to account for the osmotic pressure and HI of the interacting substrates. 
A comparison to BD
simulations excluding hydrodynamic interactions (accounted for in the
theory by $\alpha=0$) showed excellent agreement for densities up to
$B_2\rho_0\simeq0.4$ for repulsive interactions, and up to $B_2\rho_0\simeq0.20$ 
for Yukawa-like
attractions. Our BD simulations do not include HI but support our
theoretical concept and prove its validity when HI can be neglected,
i.e. for long-ranged interactions. We are confident that our
theoretical treatment is valid also in the general case
including HI; when the density-profiles of the substrates are slowly
varying in space, the correlations in the system are usually well
approximated by those in the bulk, even in the dynamic
case.\cite{dhontbook} However, a verification of our theory including
HI is highly desired and abandoned to future work, where the HI need to 
be treated by accurate means.

In principle, our study provides an alternative way to estimate the
second virial coefficient of interacting macromolecules experimentally
by measuring their steady-state rate constant in diffusion-controlled
reactions at low densities. In such experiments, the effective sink-substrate
interaction has to be known (e.g. by measuring the infinite dilution limit
of the rate constant) and the hydrodynamic quantities $D_0(r)$ and
$\alpha$ must be approximated, e.g. by using the Oseen
tensor,\cite{northrup:jcp:1984} and approximating the substrate
particles by  hard spheres with an effective hydrodynamic radius, 
respectively.\cite{dhontbook} 

The generalized Smoluchowski equation used in our work can also be
understood as a LDA in the recently proposed framework of dynamic
density functional theory (DDFT),\cite{marconi:jcp:1999} where
equilibrium correlations are used to approximate the dynamical
correlation in a Brownian system. It was shown by BD simulations that
DDFT including more sophisticated approximations than LDA works well
in the case of dense one\cite{marconi:jcp:1999} and three-dimensional
hard spheres,{\cite{royall} and three-dimensional particles with very
soft interactions,\cite{dzubiella:jpcm:2003,penna:pre:2003,archer} and
seems to provide a powerful tool to extend our work to more strongly
correlated systems, i.e nucleating or aggregating colloids,
polymerization, or binding in crowded protein solutions.\cite{zhou}

Finally, we hope that our approach will shed some light on interpretations of
experimental rate constant measurements\cite{enzyme} and will be
useful in extending existing works on the time-dependence of the rate
constant,\cite{northrup:jcp:1979,northrup:jcp:1984,vanbeijeren:jcp:2001}
crowding effects in reactions,\cite{zhou:jcp:1991,ibuki:jcp:1997} finite sink
concentration,\cite{felderhof:jcp:1976,szabo:prl:1988,gopich:jcp:2002}
or anisotropic reactivity\cite{northrup:jcp:1984,zhou:1993} to the case 
of interacting substrate particles. 

\section*{Acknowledgment}
The authors are grateful to 
Christos~N. Likos for a critical reading of the manuscript and 
Sanjib Senapati for useful
discussions. J.D. acknowledges financial support from a DFG
Forschungsstipendium. Work in the McCammon group is supported by NSF,
NIH, HHMI, CTBP, NBCR, and Accelrys, Inc.

\end{document}